\documentclass[10pt]{article}
\usepackage[utf8]{inputenc}
\usepackage[english]{babel}
\usepackage{amsfonts,amsmath,amssymb}
\usepackage{mathtools}
\usepackage{graphicx}
\usepackage[font=small,labelfont=bf]{caption}
\usepackage{booktabs}
\usepackage{wasysym}
\usepackage{multirow}
\usepackage{enumerate}
\usepackage{epstopdf}
\usepackage{bbold}
\usepackage{bbm}
\usepackage{tabularx}
\usepackage{ragged2e} 
\usepackage{booktabs} 
\usepackage{hyperref}

\newcommand{\setup}{\mathrm{setup}}

\title{Adaptive Aggregation-based\\ Domain Decomposition Multigrid\\ for Twisted Mass Fermions}
\author{Constantia Alexandrou\textsuperscript{a,b}, Simone Bacchio\textsuperscript{a, c}, 
Jacob Finkenrath\textsuperscript{b},\\ Andreas Frommer\textsuperscript{c}, Karsten Kahl\textsuperscript{c} and Matthias Rottmann\textsuperscript{c}\\
\small\textsuperscript{a}Department of Physics, University of Cyprus, PO Box 20537, 1678 Nicosia, Cyprus\\
\small\textsuperscript{b}Computation-based Science and Technology Research Center, The Cyprus Institute\\ 
\small\textsuperscript{c}Fakult\"at f\"ur Mathematik und Naturwissenschaften, Bergische Universit\"at Wuppertal}
\date{}

\begin{document}
	
\maketitle
\begin{abstract}
The Adaptive Aggregation-based Domain Decomposition Multigrid method~\cite{Frommer:2013fsa} is extended for two degenerate flavors of twisted mass fermions. By fine-tuning the parameters we
achieve a speed-up of the order of hundred times compared to the conjugate gradient algorithm for the physical value of  the pion mass. A thorough analysis 
of the aggregation parameters is presented, which provides a novel insight into multigrid methods for lattice QCD 
independently of the fermion discretization. 
\end{abstract}

\section{Introduction\label{sec:introduction}}

Lattice Quantum Chromodynamics (QCD) is computationally demanding 
due to the fact that a linear system of very large dimension needs to be 
solved repeatedly. The linear system to be solved is given by
\begin{equation} 
 D(U,\mu)\,v = b
\label{linear}
\end{equation}
where $D(U,\mu)$ is the Dirac operator. 
The Dirac operator depends on the gauge configuration $U$ and the twisted mass parameter $\mu$, which determines the quark mass.  The right hand side (rhs) $b$ is a known
spinor source and $v$ is the solution to the equation. In order to collect enough statistics for the observable under study,
Eq.~\eqref{linear} needs to be solved many times, with both varying gauge configurations and rhs. 
In fact  the so called  point-to-all quark propagator 
is obtained by solving Eq.~\eqref{linear} for twelve different rhs. 
Several of these point-to-all propagators are calculated for different space-time positions on each  gauge 
configuration and  a large number of configurations is needed to obtain sufficient statistical accuracy.  
The time for the inversion of the Dirac operator depends on   
its condition number, which gets worse as the mass of the light quarks or equivalently $\mu$ decreases.
Therefore it is crucial to use algorithms which are less sensitive to the condition number 
and adapt them to the specific features of the discretization.
This is  particularly important for inversions using gauge configurations
simulated  with u- and d- quark masses fixed to their physical value, 
where improved solvers are indispensable for speeding up the computations and 
enabling the accumulation of enough statistics.
Nowadays, such simulations are intensively being pursued by a number of lattice QCD collaborations
and  a lot of effort has been devoted in speeding up the solvers.
A very successful approach has been based on 
 multigrid ideas in preconditioning standard Krylov subspace solvers.
There is a number of variant formulations of highly optimized multigrid solvers, which yield improvements of more 
than an order of magnitude in the case of the Wilson Dirac discretization, 
as reported in Refs.~\cite{Luscher:2007es, Osborn:2010mb, Frommer:2013kla}. 
The most known implementations of multigrid algorithms, which are also available as open source software are the following: i) a two-level multigrid approach
based on L\"uschers inexact deflation~\cite{Luscher:2007se} provided in the software openQCD~\cite{Luscher:OpenQCD}; ii)
a multigrid approach developed in Refs.~\cite{Brannick:2007ue, Clark:2008nh, Babich:2009pc, Babich:2010qb}, 
referred to as MG-GCR (Multi Grid - Generalized Conjugate Residual) which is part of the USQCD package QOPQDP~\cite{USQCD:QOPQDP} and; iii) 
an adaptive aggregation-based domain decomposition multigrid approach~\cite{Frommer:2013fsa}, referred to as 
DD-$\alpha$AMG, recently made publicly available in the DDalphaAMG library~\cite{Rottmann:DDalphaAMG}. 

Although these solvers have been developed for clover Wilson fermions they can be
extended to other fermion discretization schemes as has been done for example for  the overlap operator 
using DD-$\alpha$AMG~\cite{Brannick:2014vda} or for Domain-Wall fermions using MG-GCR~\cite{Cohen:2012sh}.

In this work we focus on the twisted mass (TM) Wilson Dirac operator.  
This discretization scheme has the advantage that all observables
are automatically ${\cal O} (a)$ improved when tuned to maximal twist~\cite{Frezzotti:2003ni}, where $a$ is the lattice spacing.  This 
formulation is thus particularly suitable for hadron structure studies, since the probe such as the axial current needs no further improvement in contrast to clover improved fermions.
Furthermore, the presence of a finite twisted mass term bounds the spectrum of 
$DD^\dagger$ from below by a positive term. This avoids exceptional configurations and, at the same 
time, gives an upper bound to the condition number, improving the convergence of numerical methods used in lattice QCD. 
Adding a small twisted mass term to the 
Wilson Dirac operator is also utilized as a `trick' within simulations with the Wilson Dirac
operator. For example, it is exploited in the case of quenched calculations using Wilson
fermions~\cite{Jansen:2005kk,AbdelRehim:2005gz}, in the simulation of 
clover improved Wilson fermions where it is combined with twisted
mass reweighting~\cite{Luscher:2008tw, Luscher:2012av} or 
in mass reweighting~\cite{Finkenrath:2013soa}.
Currently, the twisted mass formulation is the discretization scheme adopted by 
the European Twisted Mass (ETM) Collaboration for studying a wide range of
observables. All these simulations employing twisted mass 
fermions will substantially benefit from an efficient multigrid method,
especially for the analysis at the  physical value of the light quark masses.

In this work we extend and optimize the DD-$\alpha$AMG method to the case of the 
twisted mass Wilson Dirac operator.
Therefore we exploit the $\Gamma_5$-preservation within DD-$\alpha$AMG
which complements the twisted mass formulation.
Our analysis shows that firstly the
optimal parameters for the method are different from those in the Wilson Dirac case,
and secondly that, with appropriate parameter choices, we can gain 
speed-ups of more than one order of magnitude
over methods previously used. The latter is similar to what has been observed in the 
case of clover improved fermions. 
Determining the  appropriate parameter set is crucial since the  tuning is a highly non-trivial task, 
and a complete analysis of how the method behaves for the  different values of the parameters is still missing in the literature.
We demonstrate in this study how the parameter space can effectively be reduced by tuning 
first the aggregation parameters. In doing so we also illustrate the robustness of the method. 
Moreover, our tuning approach can be easily applied to other aggregation-based multigrid solvers and 
fermion discretization schemes.

The paper is structured as follows: In Section~\ref{sec:twistedmass}
we introduce the twisted mass operator and discuss its properties with respect to  
multigrid preconditioning. In Section~\ref{sec:numerical} we  
present our numerical results for twisted mass fermions,
including the novel investigation of the aggregation parameters. Finally,
in Appendix~\ref{sec:library} we give an overview of the implementation of the method and how the  
DDalphaAMG library is extended to the case of twisted mass fermions~\cite{Bacchio:DDalphaAMG}.

\section{DD-$\alpha$AMG for twisted mass fermions}
\label{sec:twistedmass}

In this section we discuss the extension of the  DD-$\alpha$AMG solver to the case of the 
twisted mass fermion discretization scheme of the Dirac operator. 
For completeness we first review the twisted mass discretization scheme before  we 
 outline the various components 
of the multigrid preconditioner, pointing out the modifications 
necessary when going from the Wilson formulation to the twisted mass formulation, along the lines of Ref.~\cite{Frommer:2013fsa}.

We  work on a four-dimensional hypercubic lattice defined by
\begin{equation}
 \mathcal{V} = \{ x = (x_0,x_1,x_2,x_3), 1 \leq x_0 \leq N_T, \, 1 \leq x_1,x_2,x_3 \leq L \}
\end{equation}
with $N_T$ the number of points in the temporal direction and 
$L$ the number of points in the spatial directions $x,\,y$ and $z$.
Here and in what follows, the lattice spacing $a$ is set to unity.
The lattice volume is given by $V=N_T\cdot L^3$. Fermion fields 
are defined on the sites of the lattice, and each has 
four spin and three color degrees of freedom. The
overall space is thus
\begin{equation}
 \mathcal{V}_s = \mathcal{V} \times\mathcal{S}\times\mathcal{C}
 \label{eq:lat:sub}
\end{equation}
where  $\mathcal{S}$ denotes the spin space and $\mathcal{C}$ the color space.

\subsection{Wilson Twisted Mass fermions}\label{sec:Wtwisted}

The Wilson Dirac operator $D_W = D(m)$ with the clover term can be written as
\begin{eqnarray*}
  (D_W\psi) (x) &=& \Big( (m+4) I_{12} - \frac{c_{sw}}{32} \sum_{\mu,\nu=0}^3 ( \gamma_\mu \gamma_\nu ) \otimes \big( Q_{\mu\nu}(x) - Q_{\nu\mu}(x) \big) \Big) \psi(x) \\
             & &  \mbox{} - \frac{1}{2}\sum_{\mu=0}^3 \left( (I_4-\gamma_\mu)\otimes U_\mu(x)\right) \psi(x+\hat{\mu}) \\
             & & \mbox{} - \frac{1}{2}\sum_{\mu=0}^3 \left( (I_4+\gamma_\mu)\otimes U_\mu^\dagger(x-\hat{\mu})\right) \psi(x-\hat{\mu}) \;,
\end{eqnarray*}
with $m$ the mass parameter and  $c_{sw}$ the parameter of the clover term.
The gauge links $U_\mu(x)$ are $SU(3)$ matrices,
and the set $\{ U_\mu(x) \, : \, x \in \mathcal{L} , \, \mu=0,1,2,3 \}$ is referred to as a gauge configuration.
The $\gamma$-matrices act on the spin degrees of freedom of the spinor field $\psi(x)$
and fulfill the anti-commutation relation $\{\gamma_\mu,\gamma_\nu\} = 2 \cdot I_4 \; \delta_{\mu \nu}$
for $\mu,\nu= 0,1,2,3$.

The Wilson Dirac operator 
satisfies the relation $\Gamma_5 D_W = D_W^\dagger \Gamma_5$ referred to as $\Gamma_5$-hermiticity, where
$\Gamma_5 = I_\mathcal{V}\otimes\gamma_5\otimes I_\mathcal{C}$
acts on the space $\mathcal{V}_s$ defined in Eq.~\eqref{eq:lat:sub} as a linear
transformation of the spin degrees of freedom at each lattice site. 
In this paper we use a representation for $\gamma_5 =  \gamma_0 \gamma_1 \gamma_2 \gamma_3$,
which is diagonal in the spin space
\begin{equation}
 \gamma_5 = \begin{pmatrix} 1 & \phantom{1} & \phantom{1} & \phantom{1} \\
                               & 1           &             &             \\
                               &             & $-$1        &             \\
                               &             &             & $-$1        \\ 
  \end{pmatrix}.
\end{equation}
The clover term, $Q_{\mu\nu}$, is  given by
$$
  \begin{array}[h]{rcl}
    Q_{\mu\nu}(x) & = & U_{\mu}(x) \, U_{\nu}(x+\hat\mu) \, U_{\mu}(x+\hat\nu)^\dagger \, U_{\nu}(x)^\dagger + \\
                  &   & U_{\nu}(x) \, U_{\mu}(x-\hat\mu+\hat\nu)^\dagger \, U_{\nu}(x-\hat\mu)^\dagger \, U_{ \mu}(x-\hat\mu) + \\
                  &   & U_{\mu}(x-\hat\mu)^\dagger \, U_{\nu}(x-\hat\mu-\hat\nu)^\dagger \, U_{ \mu}(x-\hat\mu-\hat\nu) \, U_{ \nu}(x-\hat\nu) + \\
                  &   & U_{\nu}(x-\hat\nu)^\dagger \, U_{\mu}(x-\hat\nu) \, U_{ \nu}(x-\hat\nu+\hat\mu) \, U_{\mu}(x)^\dagger \; . 
  \end{array} 
$$ 
In case of the clover improved Wilson Dirac operator, the $c_{sw}$-term reduces the discretization error from $\mathcal{O}(a)$ to $\mathcal{O}(a^2)$
when  $c_{sw}$ is properly chosen. 

The twisted mass formulation is  a lattice regularization that allows automatic ${\cal {O}}(a)$ improvement
by tuning only one parameter, namely  the bare untwisted quark mass needs to be tuned to the so called
critical mass. This formulation is particularly appropriate to hadron structure studies since the renormalization 
of local operators  is significantly simplified with respect to the standard Wilson regularization. In 
the continuum, the twisted mass formulation is equivalent to the standard QCD action in a different basis. A mass term  
of the form $i\mu\tau_3 \otimes \Gamma_5 $ can be added to the standard quark mass, where $\tau_3 = \textrm{diag} (1, $$-$$1)$ is the third Pauli matrix
acting on a two-dimensional flavor space~\cite{Frezzotti:2003ni,Frezzotti:2004wz}. 
By performing a suitable  chiral transformation of the quark fields the mass term can be 
rewritten in the standard form. The twisted mass theory becomes non-trivial on  the lattice since Wilson
fermions explicitly break chiral symmetry.  We denote the twisted mass parameter by $\mu \in \mathbb{R}$ 
and apply the  twist  in the  flavor space of the up (u-) and down (d-) quark. Then the twisted mass term 
acts with a positive shift given by $i\mu \Gamma_5 $ on the u-quark operator and 
with a negative shift given by $-i\mu \Gamma_5 $ on the d-quark operator.
The twisted mass term breaks the isospin symmetry between the u- and the 
d-quark explicitly, which vanishes in the continuum limit.
In the flavor space the operator applied to a spinor field is given as
\begin{equation}
\begin{bmatrix} D(\mu) & 0 \\
                  0  & D(-\mu)  \\
  \end{bmatrix} 
  \begin{bmatrix}
  \psi_u \\
  \psi_d \\
  \end{bmatrix} =
  \begin{bmatrix}
  (D_W (\psi_u) (x)  + i\mu \Gamma_5 \psi_u (x)  \\
  (D_W (\psi_d)) (x) - i\mu \Gamma_5 \psi_d (x)
  \end{bmatrix},
\end{equation}
where
\begin{equation}
D(\mu) = D_W +i \mu \Gamma_5
\end{equation}
is what we refer to as the twisted mass Wilson Dirac operator 
defined on the space $\mathcal{V}_s$.
Adding a clover term reduces the isospin breaking induced by the twist.

Due to $\Gamma_5$-hermiticity, 
the symmetrized Wilson Dirac operator $H_W = \Gamma_5 D_W$ is hermitian
(and indefinite), such that we have
\[
H_W = V\Lambda V^\dagger,
\]
where the diagonal matrix $\Lambda$ contains the eigenvalues $\lambda_j$ of $H_W$ (which are all real)  and the unitary matrix 
$V$ the corresponding eigenvectors. The `symmetrized' twisted mass operator $H(\mu) = \Gamma_5 D(\mu) =
H_W + i  \mu $ thus satisfies
\begin{equation}
   H_W+i\mu = V \Lambda \, V^\dagger + i\mu =  V e^{i\Theta} \sqrt{\Lambda^2 + \mu^2} \; V^\dagger,
   \label{eq:Hermmu}
\end{equation}
where the diagonal matrix $e^{i\Theta}$ contains the complex phases $\theta_j$  of the eigenvalues 
$\lambda_j + i\mu$ and $\sqrt{\Lambda^2 + \mu^2}$ their absolute values. For the 
non-symmetrized twisted mass operator $D(\mu)$, analogously to the Wilson case~\cite{Frommer:2013fsa}, this gives the singular value decomposition   
\begin{equation}
   D(\mu )= \Gamma_5 V e^{i\Theta} \, \sqrt{\Lambda^2 + \mu^2}  \; V^\dagger = U \, \sqrt{\Lambda^2 + \mu^2} \; V^\dagger
\end{equation}
with $U= \Gamma_5 V e^{i\Theta}$ and $V$ being unitary. The smallest singular value $\sqrt{\lambda_i^2 + \mu^2}$ is thus not
smaller than $\mu$, which shows that a non-zero value of $\mu$ protects the twisted mass operator
$D(\mu)$ from being singular, unlike the Wilson Dirac operator where this can happen for small quark masses $m$.

Similarly, for the squared twisted mass operator 
\begin{equation}
 D^\dagger(\mu) D(\mu) = ( D^\dagger-i\mu\Gamma_5)(D+i\mu\Gamma_5)  = D^\dagger D + \mu^2 = H_W^2 + \mu^2
\label{eq:tw:sqD}
\end{equation}
we have 
\begin{equation}
 H_W^2 + \mu^2 = V (\Lambda^2 + \mu^2) V^\dagger,
\end{equation}
the eigenvalues of which are bounded from below by $\mu^2$.

Attaining automatic ${\cal O}(a)$ improvement on the lattice  can be 
accomplished  by tuning the PCAC (partial-conserved axial current) pion mass to zero. This corresponds to setting the angle $\Theta$ of the axial transformation $e^{i\Theta\tau_3 \Gamma_5}$ to $\pi/2$.
Then the renormalized light quark mass is directly proportional
to the twisted mass parameter $\mu$ with $m_R = Z_P^{-1} \mu$
and $Z_P$ the pseudoscalar renormalization constant.
The breaking of the isospin symmetry in the twisted mass formulation  results in  
the neutral pion being lighter than the charged pion. This slows
down or even prohibits simulations for light quark masses close to the physical value.  By adding the clover term to the action, 
the critical twisted mass value for the
light quarks is significantly reduced compared to simulations without this term. This allows 
simulations at the physical point with a value of the lattice spacing
around $a = 0.1$~fm or even larger~\cite{Abdel-Rehim:2015pwa}. 

\begin{figure}
\centering
\includegraphics[width=0.63\textwidth]{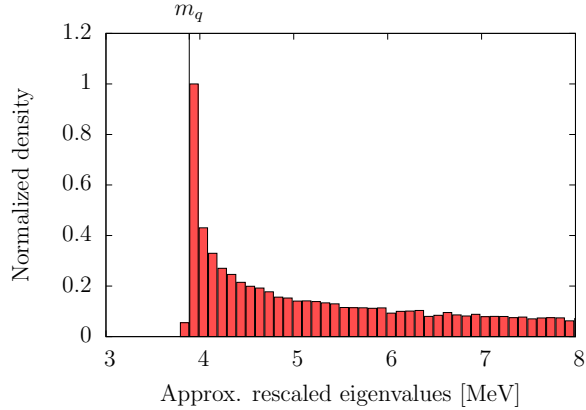}
\caption{\label{fig:evs_TM} The density of the approximated rescaled eigenvalues of the
hermitian even-odd reduced twisted mass Dirac operator measured on gauge configurations 
of the physical  ensemble $cA2.09.48$ (see  Section~\ref{sec:numerical}). 
The quark mass is given by $m_q=3.89$~MeV in the $\overline{\textrm{MS}}$-scheme.
}
\end{figure}

A special property of the twisted mass operator is that at maximal
twist the region just above $\mu^2$ is  densely populated with
the eigenvalues of the squared operator, cf.~Eq.~\eqref{eq:tw:sqD}. 
We illustrate this in Figure \ref{fig:evs_TM}, which  displays a histogram of the (scaled) ensemble averaged 
moduli of the eigenvalues of the non-squared symmetrized even-odd reduced (or preconditioned)
twisted mass Dirac operator $\hat{H}$. The eigenvalues of the operator are measured on an ensemble 
simulated at a physical value of the light quark mass doublet, which we will refer to as the
\textit{physical ensemble}. 
$\hat{H}$ is obtained using  an even-odd ordering of the lattice sites such that
\[
D= \begin{bmatrix} D_{oo} & D_{oe} \\ D_{eo} & D_{ee} \end{bmatrix}
\]
where $D_{oo}$ and $D_{ee}$ are diagonal in spinor space. $\hat{H}$ is then given  
as $\hat{H}= \Gamma_5 \hat{D}$ with
\begin{equation}
   \hat{D} = (D_{ee} - D_{eo} D^{-1}_{oo} D_{oe}).
\label{oereduced:eq}
\end{equation}

For $c_{sw}=0$, i.e.~without the clover term, the spectrum of $\hat{D}$ is directly connected
to the spectrum of the full operator $D$, e.g.~in the case of the small eigenvalues we have
\begin{equation}
 \frac{\lambda_D}{m+4} = 1 - \sqrt{1-\lambda_{\hat{D}}/(m+4)} \approx \frac{1}{2} \frac{\lambda_{\hat{D}}}{m+4}
 \label{eq:fullvseo}
\end{equation}
with $\lambda_D$ and $\lambda_{\hat{D}}$ an eigenvalue of $D$ and $\hat{D}$, respectively.
Although this relation does not hold exactly for the hermitian even-odd reduced 
twisted mass Dirac operator, we found that numerically this relation still holds approximately 
for the eigenvalues close to $\mu$.
We find that the largest relative deviation of the smallest eigenvalue
to the approximated cut-off $2 \mu$ is given by $\left|\lambda_{min}-2 \mu\right|/2 \mu<0.0005\; \textrm{MeV}$.
Thus we rescale the spectrum by a factor two and relate it to
the energy scale of the $\overline{\textrm{MS}}$-scheme defined at $2 \; \textrm{GeV}$.
The eigenvalue density is shown in Figure~\ref{fig:evs_TM}. In contrast to
the spectrum of the Wilson Dirac operator shown in Refs.~\cite{Luscher:2007se,DelDebbio:2005qa}, the density of the eigenvalues increases
close to the physical quark mass.

Since $D(\mu)$ is non-normal, the left and right eigenvectors differ. For the twisted mass Dirac operator, 
the left eigenvectors of the u-quarks, represented by $D(\mu)$,
are connected to the right eigenvectors of the d-quarks, represented by $D(-\mu)$.
If $\varphi^{u}_{j,L}$ and $\varphi^{u}_{j,R}$ are left and right eigenvectors
of $D(\mu)$, respectively, with corresponding eigenvalue $\lambda_j^{u}$, then due to $\Gamma_5$-hermiticity, we have
\begin{align}
{\lambda_j^{u}} &= {\varphi^{u}_{j,L}}^\dagger D(\mu) \varphi^{u}_{j,R} = \left( {\varphi^{u}_{j,R}}^\dagger (D(\mu))^\dagger \varphi^{u}_{j,L} \right)^\dagger = \left( {\varphi^{u}_{j,R}}^\dagger \Gamma_5 D(-\mu)  \Gamma_5 {\varphi^{u}_{j,L}} \right)^\dagger\nonumber\\
&=\left( {\varphi^{d}_{j,L}}^\dagger D(-\mu) {\varphi^{d}_{j,R}} \right)^\dagger = \mathrm{conj}(\lambda_j^{d}).
\end{align}
Thus, eigenpairs of $D(-\mu)$ are connected to the eigenpairs of $D(\mu)$ by the transformations 
$\lambda^{d}_j=\mathrm{conj}(\lambda^{u}_j)$, ${\varphi^{d}_{j,L}} = \Gamma_5 {\varphi^{u}_{j,R}}$ and ${\varphi^{d}_{j,R}} =   \Gamma_5 {\varphi^{u}_{j,L}}$.

\subsection{Multigrid preconditioning for Twisted Mass fermions\label{sec:multigrid}}

Our task is to solve the linear system
\begin{equation}\label{eq:dirac_eq}
D(\mu) \,\psi \, = \, (D+i\mu\Gamma_5) \,\psi=b
\end{equation}
for the Wilson twisted mass operator $D(\mu)$. The idea is to precondition a
flexible iterative Krylov solver at every iteration step using a multigrid preconditioner.

Let $\psi$ denote the current approximate solution to Eq.~\eqref{eq:dirac_eq}, then the 
corresponding error is given by $\epsilon = D^{-1}(\mu) b -\psi$, and
the residual $r$ satisfies
\begin{equation}
r = b-D(\mu)\psi = D(\mu)\epsilon.
\end{equation}
In lattice QCD, preconditioners like the Schwarz Alternating Procedure
(SAP~\cite{Luscher:2003qa}) efficiently reduce error components belonging to ultra-violet (UV)-modes, 
i.e.~error components belonging to eigenvectors to large eigenvalues. 
Thus the error $\epsilon$ is then dominated by infrared (IR)-modes, 
i.e.~eigenvectors corresponding to small eigenvalues.
For larger volumes, the increasing number of IR-modes slows down the preconditioned Krylov method.

Multigrid methods deal efficiently with both, IR- and UV-modes, independently of the 
volume size. Here we focus on the DD-$\alpha$AMG preconditioner.
A generic preconditioning step is described via its error propagation 
\begin{equation}\label{eq:error_propagation}
\epsilon\,\leftarrow\,\left(I-MD\right)^k\left(I-PD_c^{-1}(\mu)RD\right)\left(I-MD\right)^j\epsilon.
\end{equation}
Therein $M$ denotes the smoother, $P$ denotes the interpolation which maps from a coarser space with
less degrees of freedom to $\mathcal{V}_s$, and $R$ denotes the restriction, the counterpart of $P$, 
which maps from $\mathcal{V}_s$ to the coarser space. Having $P$ and $R$, the coarse grid operator
$D_c(\mu)$, which can be seen as a coarse version of $D(\mu)$, 
is defined by the Galerkin condition
\begin{equation}\label{eq:coarse_dirac}
D_c(\mu)=R\,D(\mu)\,P.
\end{equation}
Furthermore, $I-MD$ denotes the error propagator of the smoother. The powers $j$ and $k$ denote
it's repeated application, i.e.~we have $j$ pre-smoothing and $k$ post-smoothing steps.
This error propagator is supposed to act on the UV-modes, it reduces the error components belonging to UV-modes.
$I-PD_c^{-1}(\mu)RD$ denotes the error propagator of the coarse grid correction which acts 
complementary to the error propagator of the smoother. It reduces the error components belonging to IR-modes.
As an iteration prescription, the coarse grid correction itself is given by
\begin{equation}\label{eq:multigrid_preconditioning}
\psi\,\leftarrow\,\psi+PD_c^{-1}(\mu)R\,r.
\end{equation}
This two-level approach can be recursively expanded to a multi-level approach by using again a two-level approach
for the inversion of $D_c(\mu)$.
Then on each new level $\ell$, transfer operators
$R_\ell$ and $P_\ell$  
are employed in the construction of the next level operator $D_{\ell+1}(\mu)=R_\ell D_\ell(\mu)P_\ell$
generalizing Eq.~\eqref{eq:coarse_dirac}, and a smoother $M_\ell$ is assigned.
DD-$\alpha$AMG uses a K-cycle approach~\cite{Notay2007}, which means that at each level,
a multigrid preconditioned Krylov method is performed until a prescribed decrease of the residual is obtained.
Our numerical tests showed that the inversion of the coarsest Dirac operator $D_L$ can be evaluated at low accuracy, 
which reduces the computational effort. 
This is not the case for the inexact deflation approach from Ref.~\cite{Luscher:2007se} as 
explained in Ref.~\cite{Frommer:2013fsa}.

\subsection{Coarse grid}\label{sec:cgrid}
The coarse grid correction, given in Eq.~\eqref{eq:multigrid_preconditioning},
should approximately remove the error corresponding to the IR-mode components on the fine grid.
An ideal, yet naive, choice would be to span the prolongation 
operator $P$ by the right eigenvectors corresponding to small eigenvalues of $D(\mu)$
and the restriction operator $R$ by the left ones,
since then the coarse grid correction would entirely remove the (right) IR-modes present in the error.
Such a choice is very similar to exact deflation of the 
small eigenvalues in that the coarse grid correction would solve directly the linear system of the IR-modes.
In practice, the computational effort for generating such an ``exact`` coarse grid projection
is more than a magnitude higher than in aggregation based approaches, 
see e.g.~Table~\ref{tab:clover} and~\ref{tab:twistedmass} in Section \ref{sec:compare_CG},
and it scales with $\mathcal{O}(V^2)$. Instead,  the DD-$\alpha$AMG solver uses a coarse grid 
correction, which is based on aggregation and the property of \textit{local coherence} described in Ref.~\cite{Luscher:2007se}. 
Local coherence implies  that it is possible to approximate the subspace of the IR-modes by aggregates over
just a small set of $\mathcal{O}(20)$ {\em test vectors} $v_i$. 
In this section, we discuss how to construct the aggregation based prolongation, the restriction and the coarse grid operators 
from the test vectors, while in Section~\ref{sec:setup} we will describe the
strategy for generating the test vectors.

\begin{figure}
\centering
\includegraphics[width=0.7\textwidth]{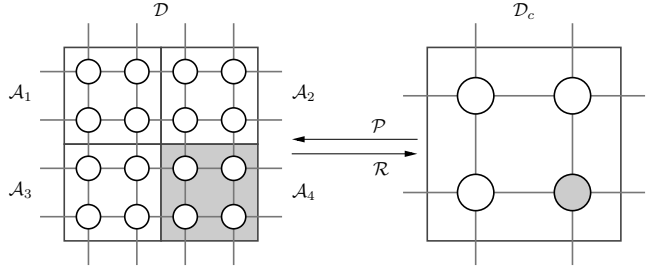}
\caption{\label{fig:aggregation} 
 Block aggregation of the Dirac operator and symbolic representation
of the movement from the fine operator to the coarse operator and vice-versa.}
\end{figure}

The coarse grid is obtained by mapping an aggregate $\mathcal{A}_j$ of the fine grid 
to a single site of the coarse grid, where
\begin{equation}\label{eq:aggregate}
\mathcal{A}_j = \mathcal{V}_{j} \times \mathcal{S}_j \times \mathcal{C}_j.
\end{equation}
We will use static blocks with a fixed size given by
\begin{equation}
\mathcal{V}_j = \mathcal{T}_{j} \times \mathcal{Z}_{j} \times \mathcal{Y}_{j} \times \mathcal{X}_{j}
\label{eq:blocks}
\end{equation}
such that these blocks decompose the lattice $\mathcal{V}_s$ 
as shown in Figure \ref{fig:aggregation}.
The number of sites on the coarse lattice $\mathcal{V}_c$ is then given 
by $N_b=V/V_b$ with $V_b$ the block volume.

Now, a projection $P$ and a restriction $R$ between the fine grid and the coarse grid 
\begin{equation}
 \mathcal{V} \times \mathcal{S} \times \mathcal{C}\quad\xrightleftharpoons[~~P~~]{~~R~~}\quad\mathcal{V}_c \times N_v,
\end{equation} 
can be constructed by using the test vectors $v_i$. For the prolongation $P$
the test vectors are decomposed into blocks over the aggregates as 
\begin{equation}\label{eq:prolong}
  (v_1 \mid \ldots \mid v_{N_v}) = 
  \left( \begin{array}{|c|}
      \hline
      \cline{1-1}
      \multicolumn{1}{||||||}{} \\
      \multicolumn{1}{||||||}{} \\
      \multicolumn{1}{||||||}{} \\
      \multicolumn{1}{||||||}{} \\
      \multicolumn{1}{||||||}{} \\
      \multicolumn{1}{||||||}{} \\
      \multicolumn{1}{||||||}{} \\
      \hline
    \end{array} \right)
  \;\longrightarrow\;
  P = \left( \begin{array}{c c c c c c c c}
      \cline{1-1}
      \multicolumn{1}{||||||}{} & & & \\
      \multicolumn{1}{||||||}{} & & & \\
      \cline{1-2}
      & \multicolumn{1}{||||||}{} & & \\
      & \multicolumn{1}{||||||}{} & & \\
      \cline{2-2}
      & & \ddots & \\
      \cline{4-4}
      & & & \multicolumn{1}{||||||}{} \\
      & & & \multicolumn{1}{||||||}{} \\
      \cline{4-4}
    \end{array} \right)
  \begin{array}{c}
    \multirow{2}{*}{$\mathcal{A}_1$} \\ 
    \\
    \multirow{2}{*}{$\mathcal{A}_2$} \\
    \\
    \vdots \\
    \multirow{2}{*}{$\mathcal{A}_{N_b}$} \\   
    \\
  \end{array}
  .
\end{equation}
For the restriction operator, $R^\dagger$ is constructed similarly
from a possibly different set of test vectors $w_i$. For numerical reasons,
the parts of all test vectors over a given aggregate are orthonormalized. 

It is possible to extend the $\Gamma_5$-hermiticity of the Wilson Dirac operator
to the coarse grid operator. This can be done by decomposing the aggregation in the spin space $\mathcal{S}$
with an aggregate
\begin{equation}
\mathcal{A}_{j,+} = \mathcal{V}_{j} \times \mathcal{S}_{0,1} \times \mathcal{C},
\end{equation}
which collects the two upper spin components $0$ and $1$,
and an aggregate
\begin{equation}
\mathcal{A}_{j,-} = \mathcal{V}_{j} \times \mathcal{S}_{2,3} \times \mathcal{C},
\end{equation}
which collects the two lower spin components $2,3$. In the chiral basis,
$\Gamma_5$ acts with $+1$ to the components of the aggregate $\mathcal{A}_{j,+}$
while with $-1$ on the other aggregates. Now, the coarse grid $\Gamma_{5,c}$ can be defined by
$\Gamma_{5,c} = I_{\mathcal{V}_c}\otimes\tau_3\otimes I_{N_v}$,
where $\tau_3$ acts on the different spin aggregates. 
This type of aggregation was proposed 
in Ref.~\cite{Clark:2008nh} for the MG-GCR method and 
it is used as well in the DD-$\alpha$AMG method, where it is termed \textit{standard aggregation}.
The $\Gamma_5$-compatible prolongation $P$ satisfies
\begin{equation}\label{eq:gamma5_symmetry}
\Gamma_5\,P = P\,\Gamma_{5,c}
\end{equation}
which can be represented as
\begin{equation}
\resizebox{0.9\linewidth}{!}{$
\left( \begin{array}{c c c c c c c}
	\multicolumn{2}{c}{\multirow{2}{*}{$I_{6\cdot V_b}$}} &  &                                                                                &  &                                            &  \\
	 &                                                                               &  &                                                                                &  &                                            &  \\
	 &                                                                               & \multicolumn{2}{c}{\multirow{2}{*}{$-I_{6\cdot V_b}$}} &  &  \\
	 &                                                                               &  &                                                                                &  &                                            &  \\
	 &                                                                               &  &                                                                                & \multicolumn{2}{c}{\multirow{2}{*}{$\ddots$}} &  \\
	 &                                                                               &  &                                                                                &  &                                            &
\end{array} \right)
\left( \begin{array}{c c c}
	\cline{1-1}
	\multicolumn{1}{||||||}{} &                           &  \\
	\multicolumn{1}{||||||}{} &                           &  \\ \cline{1-2}
	                          & \multicolumn{1}{||||||}{} &  \\
	                          & \multicolumn{1}{||||||}{} &  \\ \cline{2-2}
	                          &                           & \multirow{2}{*}{$\ddots$} \\
	                          &                           &
\end{array} \right)
=
\left( \begin{array}{c c c}
	\cline{1-1}
	\multicolumn{1}{||||||}{} &                           &  \\
	\multicolumn{1}{||||||}{} &                           &  \\ \cline{1-2}
	                          & \multicolumn{1}{||||||}{} &  \\
	                          & \multicolumn{1}{||||||}{} &  \\ \cline{2-2}
	                          &                           & \multirow{2}{*}{$\ddots$} \\
	                          &                           &
\end{array} \right)
\scalebox{0.5}[1]{$\left( \begin{array}{c c c}
	\scalebox{2}[1]{$I_{N_v}$} &                                          &  \\
	                                        & -\scalebox{2}[1]{$I_{N_v}$} &  \\
	                                        &                                          & \scalebox{2}[1]{$\ddots$}
\end{array} \right)$}
$}.
\end{equation}
With the standard aggregation the coarse grid operator with a finite twisted mass term is given by
\begin{equation}
 D_c(\mu) =R\,\left(D + i\mu \Gamma_5 \right)\,P =  R\,D\,P + i\mu \Gamma_{5,c}  RP.
\end{equation}
As  pointed out in Ref.~\cite{Frommer:2013fsa}, the natural choice for the restriction operator is
$R= (\Gamma_5 P)^\dagger$ since in this way the restriction approximates the subspace
of the small {\em right} eigenvectors of the Wilson Dirac operator for the case 
that the interpolation approximates the small {\em left} eigenvectors.
Using this formulation, the resulting coarse grid operator $RD(\mu)P$ is equivalent to $H_{c}+ i\mu$
with $H_c = P^\dagger H P$, which is a complex shifted maximally indefinite operator (cf.~Eq.~\eqref{eq:Hermmu}).
Due to the commutativity relation between $P$ and $\Gamma_{5}$, given in Eq.~\eqref{eq:gamma5_symmetry}, and already  
noted in Ref.~\cite{Babich:2009pc}, the coarse grid corrections obtained by 
$R = (\Gamma_{5}P)^{\dagger}$ and $R = P^{\dagger}$ are identical.  
DD-$\alpha$AMG thus uses $R = P^{\dagger}$, and the coarse grid twisted mass operator their is defined by
\begin{equation}
 D_c(\mu) = P^\dagger \,D\,P + i\mu \Gamma_{5,c}.
\end{equation}
$D_c(\mu)$ preserves several important properties of the fine grid operator $D(\mu)$. 
It preserves the sparse structure in that only neighboring aggregates are coupled.
In the square coarse grid operator 
\begin{equation}
D_c^\dagger(\mu)  D_c(\mu) = D_c^\dagger D_c  + \mu^2,
\end{equation}
the eigenvalues are again bounded from below by $\mu^2$, and there is a $\Gamma_{5,c}$-symmetry
which reproduces the connection between the u- and d-quark operators as
\begin{equation}
D_{c}^\dagger(\mu) = P^\dagger (D + i \mu \Gamma_5)^\dagger P = P^\dagger (\Gamma_5 D \Gamma_5 - i \mu \Gamma_5) P =\Gamma_{5,c}\,D_{c}(-\mu) \Gamma_{5,c}\,,
\end{equation} 

Although the dimension of the coarse grid operator is reduced, it can
develop a large number of small eigenvalues close to $\mu$. This can critically slow down
the convergence of a standard Krylov solver to be used on the coarsest grid such that the time spend 
in the coarsest operator inversions dominates by far the overall inversion time even though only poor accuracy
is required. We therefore decrease the density of small eigenvalues of the coarsest grid operator
by increasing the twisted mass parameter by a factor of $\delta$ given by
\begin{equation}
 D_c(\mu, \delta) = D_c + i \delta \mu  \cdot \Gamma_{5,c}
 \label{delta:eq}
\end{equation}
with $\delta \geq 1$\footnote{This trick is used in heavy $N_f=2$ twisted mass simulations at the charm quark mass and 
we thank Bj\"orn Leder for this suggestion.}.

We will analyze the effect of $\delta$ in detail in Section~\ref{sec:trick}. As compared to the standard Wilson 
Dirac operator, for the twisted mass Dirac operator it turns out that the overall execution time is minimized if one relaxes 
even further  the accuracy of the coarsest grid solve as will be described in more detail in Section~\ref{sec:trick} as well.

\subsection{Smoother}\label{sec:smother}
In the DD-$\alpha$AMG approach, a red-black Schwarz Alternating Procedure (SAP) is used as a smoother. 
This domain decomposition method was introduced to lattice QCD 
in Ref.~\cite{Luscher:2003qa}, where it was used as a preconditioner.
The lattice is partitioned into alternated ``red(r)'' and ``black(b)'' lattice 
blocks in a checkerboard manner, and the subdomains are obtained as the 
full color-spin space over the respective lattice block (see Eq.~\eqref{eq:blocks}).
Re-ordering the Dirac twisted mass operator
such that the red blocks come first, we obtain
\begin{equation}\label{eq:D_reb_black}
D + i \mu \Gamma_5  =\left( \begin{array}{c c}
D_{rr} + i \mu \Gamma^r_5 & D_{rb}\\
D_{br} & D_{bb} + i \mu \Gamma^b_5
\end{array} \right),
\end{equation}
where $D_{rr}$ and $D_{bb}$ are block diagonal matrices filled with the respective subdomains,
while $D_{rb}$ and $D_{br}$ connect the neighboring blocks.
Note that the operator $D$ has only next neighbor interactions and thus blocks of a
specific color do not couple to the same color. The eigenmodes of the blocks have a 
higher cut-off than the full operator, given by $p_\nu = \pi/L_b$ in the free case 
for $\mu=0$ due to the Dirichlet boundary conditions.  

If now the role of the smoother is to reduce the UV-modes, a natural choice for the operator 
$M$ in the error propagation (see \eqref{eq:error_propagation}) is given by the inverse of the block operator,
resulting in
\begin{equation}\label{eq:M_SAP}
E_{\mathrm{SAP}}=\left(I-M_{\mathrm{SAP}}D(\mu)\right)=\left(I-B_{rr}(\mu)D(\mu)\right)\left(I-B_{bb}(\mu)D(\mu)\right),
\end{equation}
where $B_{rr}(\mu)$ and $B_{bb}(\mu)$ are the block inverses defined as
\begin{equation}\label{eq:B_reb_black}
B_{rr}=\left( \begin{array}{c c}
(D_{rr} + i \mu \Gamma^r_5)^{-1} & 0\\
0 & 0
\end{array} \right) \quad\text{and}\quad 
B_{bb}=\left( \begin{array}{c c}
0 & 0\\
0 & (D_{bb}+ i \mu \Gamma^b_5)^{-1}
\end{array} \right).
\end{equation}
In practice, the (approximate) inversion of the blocks on the diagonal of $B_{rr}$ and $B_{bb}$ is performed with small 
computational cost by a few steps of an iterative method like the Minimal Residual method~\cite{Frommer:2013fsa, Saad:1984:CAT}.
Note that we fix the block size to coincide with the aggregates on each level of the multigrid hierachy.

\subsection{Krylov subspace methods}\label{sec:solver}
A Krylov subspace method preconditioned by the chosen multigrid approach has to be a 
flexible algorithm, since the smoother as well as the solver on the coarse grid system are non-stationary processes. 
Flexible solvers, which have been employed in multigrid 
preconditioning for lattice QCD are the flexible BiConjugate Gradient Stabilized method (BiCGStab,~\cite{Vogel2007}),
Generalized Conjugate Residual (GCR,~\cite{Eisenstatetal1983}) and the Flexible Generalized Minimal
RESidual ((F)GMRES,~\cite{Saad1993}) solver. In the DD-$\alpha$AMG approach a FGMRES solver is used 
for the inversion of the fine grid operator and in the K-cycle for the inversion of the coarser operators
except for the coarsest. The latter is inverted by even-odd preconditioned GMRES,
i.e.~GMRES is run for the even-odd reduced system
$\hat{D}\phi_o = \eta_o-D_{oe}D_{ee}^{-1}\eta_e$ with $\hat{D}$ from Eq.~\eqref{oereduced:eq} and then $\phi_e$
in the solution $[\begin{smallmatrix}
\phi_e \\ \phi_o \end{smallmatrix}] $
of $D [ \begin{smallmatrix} \phi_e \\ \phi_o \end{smallmatrix} ] =  [\begin{smallmatrix} \eta_e &\\\eta_o \end{smallmatrix}]$
is retrieved as $\phi_e = D_{ee}^{-1}(\eta_e - D_{eo}\phi_e).$
Even-odd preconditioning is also used in the smoother when inverting the blocks. In both cases
a speed-up of up to $50\%$ compared to the full operator can be achieved.
This can be explained by the fact that the small eigenvalues are increased by a factor of two as 
seen from Eq.~\eqref{eq:fullvseo}.

In general, the accuracy of the coarse grid inversions may be very much relaxed as compared to
the target accuracy of the fine grid inversion.
Indeed, for DD-$\alpha$AMG with the K-cycle strategy, optimal results are obtained when requiring the 
(approximate) inversions of the coarser operators to reduce the residual by just one order of magnitude.

\subsection{Setup phase\label{sec:setup}} 
In the setup phase we have to compute a sufficient number of approximate low modes $v_i$, which when
chopped into aggregates, will approximate the IR-modes well due to the local coherence property.
We employ the setup algorithm from Ref.~\cite{Frommer:2013fsa} for the twisted mass operator.

The vectors $\{v_i\}$ are generated by using a variant of block inverse iteration
\begin{equation}
 v_i^{(k)} =  D^{-1}(\mu)  v_i^{(k-1)}, \, i=1,\ldots, N_v, \, k=1,\ldots,n_{\setup}, v_i^{(0)} \mbox{chosen randomly},  
\end{equation}
where the $v_i^{(k)}$ vectors converge to the eigenvectors with eigenvalues of smallest modulus.
In practice, in order to maintain numerical stability, after each iteration $k$ the vectors $v_i^{(k)}$ spanning
the space of approximate IR-modes have to be orthonormalized, and $D^{-1}(\mu)$ is replaced by a multigrid iteration
with prolongation operator constructed from the current set of approximate low modes $v_i^{(k)}$.
For $k=1$, where a multigrid hierarchy is not yet available, we just apply some steps of the SAP smoother.
This approach results in a self-adapting procedure where the multigrid hierarchy is improved while using it to 
expose the small eigenmodes. Typically, a small number of setup iterations $n_{\setup}$ is sufficient.

\section{Numerical results for twisted mass fermions\label{sec:numerical}}
The DD-$\alpha$AMG approach uses a wide range of parameters, and its time to solution
will depend on a good choice of the parameters. The purpose of this section is therefore threefold. 
We first provide a set of default parameter choices, which in our extensive numerical testing turned out to yield good 
overall performance. Secondly, we show that for physically relevant configurations, appropriately chosen parameters yield 
speed-ups of about two orders of magnitude compared to standard methods for twisted mass fermions. We also show numerically that these 
parameters can be kept fixed over a whole statistical ensemble without a notable decrease in performance. Finally, we present an 
analysis on the dependence of several parameters which implies a general strategy for obtaining good parameters for a given ensemble and computer. 
All the presented numerical results have been obtained on SuperMUC phase 2 at the Leibniz Supercomputing Centre,
an Intel Haswell Xeon computer on which we used up to 4096 cores. We have also performed 
runs on JURECA at the Julich Supercomputing Centre and on Cy-Tera at the Computation-based Science and Technology Research Center  Cyprus  obtaining compatible results.
The stopping criterion of the overall iteration was fixed such that the residual is reduced by a factor of $10^9$.

\subsection{Default parameters\label{sec:param_tune}}
Table~\ref{tab:TMparams} summarizes our default parameter set used for DD-$\alpha$AMG.
The parameter tuning was done 
for the ensemble $cA2.09.48$~\cite{Abdel-Rehim:2015pwa} 
with a lattice volume of $V=96\times48^3$ and lattice spacing of 
$a=0.0931(10)\;\textrm{fm}$ as determined from the nucleon mass. 
This ensemble was generated at a pion mass close to the physical one, namely $m_\pi=0.131$~MeV~\cite{Abdel-Rehim:2015pwa}.

\begin{table}
\centering\resizebox{\textwidth}{!}{
\begin{tabular}{llcc}
	\toprule
	          & parameter                                                        &             &   optimal   \\ \midrule
	Multigrid & number of levels                                                 & $n_{\ell}$  &     $3$     \\
	          & number of setup iterations                                       & $n_{\setup}$ &     $5$     \\
	          & number of test vectors on level $1$                              &  $N_{v,1}$  &    $28$     \\
	          & number of test vectors on level $2$                              &  $N_{v,2}$  &    $28$     \\
	          & size of lattice-blocks for aggregates on level $1$               &  $V_{b,1}$  &    $4^4$    \\
	          & size of lattice-blocks for aggregates on level $\ell, \; \ell>1$ &  $V_{b,2}$  &    $2^4$    \\ \midrule
	Solver    & mixed precision FGMRES                                           &             &  \\
	          & relative residual tolerance (restarting criterion)               &             &  $10^{-6}$  \\ \midrule
	Smoother  & red-black multiplicative SAP                                     &             &  \\
	(SAP)     & size of lattice-blocks on level 1                                &             &    $4^4$    \\
	          & size of lattice-blocks on level $\ell, \; \ell>1$                &             &    $2^4$    \\
	          & number of post-smoothing steps                                   &             &     $4$     \\
	          & MINRES iterations to invert the blocks                           &             &     $3$     \\ \midrule
	K-cycle   & with single precision FGMRES                                     &             &  \\
	          & restart length                                                   &             &     $5$     \\
	          & number of maximal restarts                                       &             &     $2$     \\
	          & relative residual tolerance (stopping criterion)                 &             &  $10^{-1}$  \\ \midrule
	Coarsest  & solved by even-odd preconditioned  GMRES                         &             &  \\
	~grid     & twisted mass parameter                                           &   $\mu_{\rm coarse}$   & $5.2\cdot\mu$ \\
	          & restart length                                                   &             &    $100$    \\
	          & number of maximal restarts                                       &             &     $5$     \\
	          & relative residual tolerance (stopping criterion)                 &             &  $10^{-1}$  \\ \bottomrule
\end{tabular}}
\caption{ The parameter set used in DD-$\alpha$AMG, obtained by parameter tuning for the 
TM fermion ensemble $cA2.09.48$~\cite{Abdel-Rehim:2015pwa}.
}
\label{tab:TMparams}
\end{table}

\subsubsection{Discussion}

\begin{table}
\resizebox{\textwidth}{!}{
	\begin{tabular}{l c c c c}
		\toprule
		\multirow{2}{*}{Solver} & Setup time    & Inversion time & Total iteration count & Total iteration count     \\
		& [core-hrs]    & [core-hrs]     & of the fine grid solve  & of coarse grid solvers \\ \midrule
		CG                      & $\textendash$ & 174.8          & 26\,937            & $\textendash$           \\ \midrule
		CG-ede                   & 1\,527.4      & 5.4            & 649                & $\textendash$           \\ \midrule
		DD-$\alpha$AMG          & 13.3          & 0.9            & 16                 & 2\,988                  \\ \bottomrule
	\end{tabular}}
\caption{\label{tab:clover} Results for clover Wilson fermions${}^\dagger$. Comparison of CG (tmLQCD), 
	eigCG (tmLQCD+ARPACK) with 800 eigenvectors and DD-$\alpha$AMG with parameters from Ref.~\cite{Frommer:2013kla}.
	The results were for a $48^3\times64$ lattice from ensemble $VII$ of Ref.~\cite{Bali:2014nma} with $m_\pi = 0.1597(15)$~GeV.}
\end{table}
\renewcommand*{\thefootnote}{\fnsymbol{footnote}}
\footnotetext[2]{In Tables~\ref{tab:clover} and~\ref{tab:twistedmass} the timings for CG and eigCG have been normalized to
1.0 Gflop/s per core (average of standard performance $\sim0.7$Gflop/s and optimal performance 
$\sim 1.3$Gflop/s); the rhs of the equation $D\psi=b$ has been randomly generated and in
all the cases the propagator $\psi$ has been computed to a relative precision of $10^{-9}$. }
\renewcommand*{\thefootnote}{\arabic{footnote}}

As explained in Section \ref{sec:setup}, in each iteration of the adaptive setup routine 
the currently available multigrid hierarchy is used to perform one iteration
with the multigrid preconditioner on each test vector.  If not stated otherwise,
we used the parameters from Table \ref{tab:TMparams} for the preconditioner in the setup phase
and in the solve phase.

As the setup iteration proceeds the test vectors become more rich in low mode components.
This also leeds to a more ill-conditioned coarse operator $D_c$ and higher iteration counts 
on the coarse grid (cf.~Ref.~\cite{Frommer:2013fsa}), i.e.~one can observe a higher cost per setup iteration as the
setup proceeds. This can be seen in in Table \ref{tab:twistedmass} where the setup times for $n_{\setup} = 3$ 
and $n_{\setup} = 5$ are stated. Indeed 
a factor of 5 in compute time between both setups can be observed. The suggested value
$n_{\setup}=5$ from Table \ref{tab:TMparams} thus yields a good value when several inversions ($\mathcal{O}(100)$ and more) 
with the same operator are desired. The relatively large setup time can be neglected 
in this case. We did not find that more setup iterations than $n_{\setup} = 5$ yield 
substantial further improvement in the solve time. On the other hand, when solving 
for a few rhs, a good balance of setup and solve time has to be found. 
Therefore a smaller number of setup iterations like $n_{\setup} = 3$ might be more suitable.

For larger pion masses one can, in principle, improve
the generation of the subspace of small eigenmodes in the setup by using a smaller 
mass parameter $m$ or a smaller twisted mass parameter $\mu$, which helps to probe
the small eigenmodes more rapidly within the setup phase.  
However, at least at the physical point, we do not find a significant improvement
by using different mass parameters.

We use the same subspace for the u-quark with $+\mu$ and the d-quark  with $-\mu$,
i.e.~we run the setup phase only once for both quarks. Although the eigenspace changes, 
numerically  we do not find  a large difference. This also saves computing time 
for many applications where the square operator has to be inverted, e.g.~during the 
integration within the Hybrid Monte Carlo (HMC) algorithm~\cite{Duane:1987de,Gottlieb:1987mq}.

\subsubsection{Comparison with CG\label{sec:compare_CG}}

To put the improvements for twisted mass simulations in perspective, we start with
an experiment for Wilson Dirac fermions, thus complementing the results from Ref.~\cite{Frommer:2013kla}
for our target machine. We use the default
parameter set from Ref.~\cite{Frommer:2013kla} and a configuration 
from ensemble $VII$~\cite{Bali:2014nma} with a pion mass of $a m_\pi = 0.05786$,
a lattice spacing of $a = 0.0071 \; \textrm{fm}$ and a lattice volume
of $V=64\times48^3$. 

Table~\ref{tab:clover} presents a comparison 
of the inversion times of the Conjugate Gradient (CG) solver, the CG solver with exact deflation 
(CG-eDe)~\cite{Saad:1984:CAT,Neff:2001zr} and the DD-$\alpha$AMG solver.
For  CG and CG-eDe we use the publicly available software package \textit{tmLQCD}~\cite{Jansen:2009xp} that
is commonly used for simulations with the twisted mass operator and 
provides a count of floating point operations per second ($Flop/s$) for the CG solver.
This makes it possible to rescale results obtained on different systems.

The conjugate gradient solver requires a positive definite hermitian matrix,
which is obtained by solving the linear system with the squared even-odd reduced operator given 
by $\hat{D}^\dagger \hat{D} x = b'$ with $b'=\hat{D}^\dagger b$.
This squares the condition number of the involved matrix. 

The CG-eDe and the DD-$\alpha$AMG solver involve a setup phase, which has to be done 
for each new configuration once before the linear system is solved.
In the case of the CG-eDe solver, $\mathcal{O}(100)$ lowest eigenmodes
of the squared even-odd reduced Dirac operator are calculated 
(here we use 800 for the case of Wilson case and 1600 for the twisted mass case).
The number of eigenvectors in CG-eDe is obtained by optimizing the time to solution (setup + inversion time)
for computing $\mathcal{O}(1000)$ rhs. Indeed the setup phase is extremely expensive 
which makes CG-eDe inefficient for a small number of rhs.
The low mode computation is done by using the publicly available package ARPACK 
with tmLQCD~\cite{Jansen:2009xp}.

\begin{table}
\resizebox{\textwidth}{!}{
\begin{tabular}{l c c c c}
	\toprule
	\multirow{2}{*}{Solver} & Setup time            & Inversion time        & Total iteration count  & Total iteration count        \\
	                        & [core-hrs]            & [core-hrs]            & of the fine grid solve & of coarse grid solvers   \\ \midrule
	CG                      & $\textendash$         & 338.6                 & 34\,790             & $\textendash$             \\ \midrule
	CG-eDe                   & 6\,941.1              & 9.8                   & 695                 & $\textendash$             \\ \midrule
	DD-$\alpha$AMG          & \multirow{2}{*}{7.7} & \multirow{2}{*}{2.5} & \multirow{2}{*}{28} & \multirow{2}{*}{16\,619} \\
	for $n_{\setup} = 3$     &                       &                       &                     &  \\ \midrule
	DD-$\alpha$AMG          & \multirow{2}{*}{38.3} & \multirow{2}{*}{1.5}  & \multirow{2}{*}{15} & \multirow{2}{*}{11\,574}  \\
	for $n_{\setup} = 5$     &                       &                       &                     &  \\ \bottomrule
\end{tabular}}
\caption{\label{tab:twistedmass} Results for TM fermions${}^\dagger$. Comparison of CG (tmLQCD), 
eigCG (tmLQCD+ARPACK) with 1600 eigenvectors and DD-$\alpha$AMG. 
The results were computed for the  $cA2.09.48$ ensemble~\cite{Abdel-Rehim:2015pwa} with $m_\pi = 0.131$~GeV.}
\end{table}

Table~\ref{tab:clover} shows that in the case of Wilson Dirac fermions, the DD-$\alpha$AMG solver 
speeds up the time to solution by roughly a factor of 200 compared to CG and roughly by a factor 
five compared to the CG-eDe solver. When including the setup time, DD-$\alpha$AMG is roughly a factor of
12 faster than CG for one rhs,
while CG-eDe is not competitive due to its computationally demanding setup.

In the twisted mass fermion case, we are able to achieve the same speed-ups as 
for the Wilson fermion case. This it not straight forward as it requires the coarsest grid
twisted mass $\mu_{\rm coarse}$ to be chosen different from the fine grid twisted mass $\mu$.
We choose $\mu_c = 5.2 \mu$ (cf.~Table \ref{tab:TMparams}) for reasons that will be explained in Section \ref{sec:trick}.
The results for twisted mass fermions are shown in Table~\ref{tab:twistedmass}. 

Indeed, we are able to achieve a speed-up in inversion time of roughly a factor of 220 
compared to CG and roughly by a factor six compared to the CG-eDe solver, with the time for the setup
being almost a factor of 100 less as compared to  CG-eDe. 
For $n_{\setup} = 3$ we used the parameters from Table~\ref{tab:TMparams} 
except for the number of test vectors on the fine level $N_{v,1}$ being 20 instead of 28.
For $n_{\setup} = 5$ we used a further optimized set which is given in the last row of Table \ref{tab:blk_cmu}.
This yields another factor of 1.5 speed-up in the inversion time at the expense of increasing the setup time.

\begin{figure}
\includegraphics[width=\textwidth]{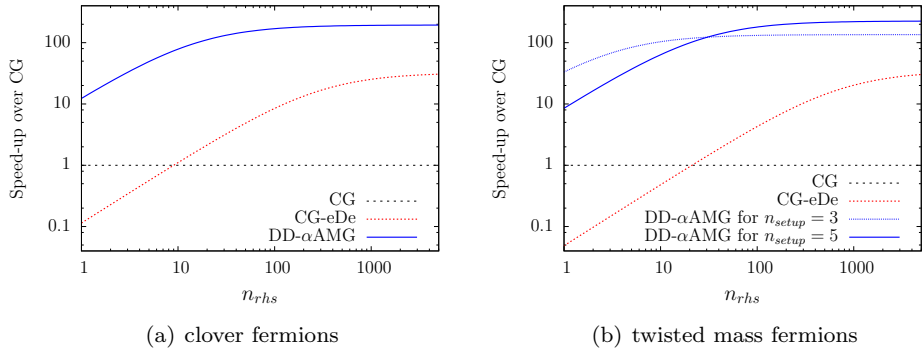}
\caption{\label{fig:comparisonCG}Speed-up over CG using the results of Table~\ref{tab:clover} and Table~\ref{tab:twistedmass}.}
\end{figure}

In Figure~\ref{fig:comparisonCG} we show the speed-up in total time 
(setup+solve) compared to one CG solve for DD-$\alpha$AMG 
and CG-eDe as a function of the number of rhs $n_{rhs}$.
The difference between the two blue curves in the figure on the right hand side 
is due to results from a different number of  setup iterations (3 and 5) 
and a different number of test vectors (20 and 28), where 3 setup iterations
are optimal for few rhs (blue dotted line) and 5 setup iterations 
for many rhs (blue line).
In summary, the results for twisted mass fermions show that for one rhs 
DD-$\alpha$AMG is roughly 
30 times faster than CG for one rhs, 120 times faster for twenty rhs and 220 times
faster for a thousand rhs.

\subsection{Stability of optimal parameters}
\begin{figure}
	\includegraphics[width=1\textwidth]{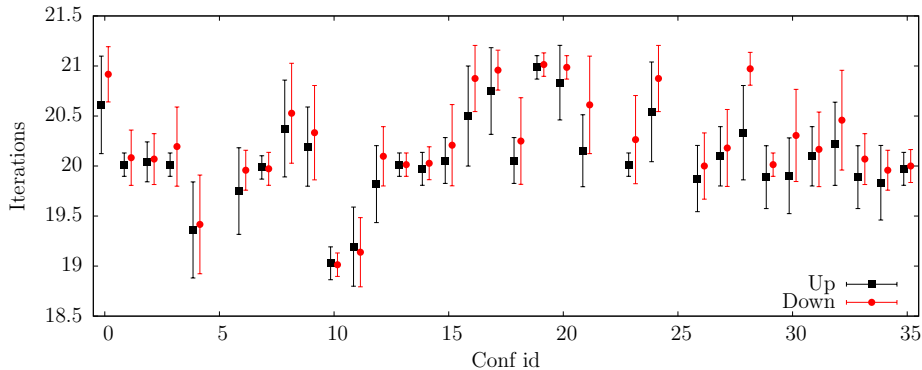}
	\caption{\label{fig:observable}\label{fig:stability} We depict the average iteration
    counts on computing quark propagators for several configuration of the ensemble $cA2.09.48$.
	For the black square points the setup is generated with the same $\mu$ while
	for the red circle points the setup is generated with a TM parameter $\mu$ with opposite sign.}
\end{figure}

\begin{figure}
\centering
\includegraphics[width=0.6\textwidth]{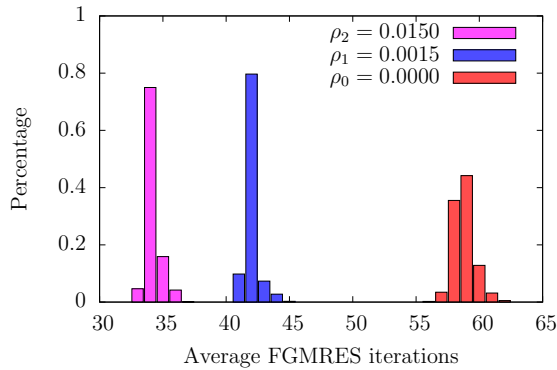}
\caption{\label{fig:hmc}We show the FGMRES iteration counts averaged over the trajectory when the solver 
is used for computing the force terms in an HMC simulation.
The data is for the ensemble $cA2.09.64$ with statistics of 2000 trajectories and the squared
operator $\hat D^\dagger \hat D + \rho_i^2$ is inverted, i.e.~two inversions are performed with 
DD-$\alpha$AMG inverting $\hat D + i \rho_i \hat\Gamma_5$ and $\hat D^\dagger - i \rho_i \hat\Gamma_5$ 
where $\hat\Gamma_5$ is the restriction of $\Gamma_5$  to the odd lattice sites. 
The parameter $\rho_i$ sets the mass for the Hasenbusch preconditioning~\cite{Hasenbusch:2002ai}, 
the integration scheme is equivalent to the one presented in Ref.~\cite{Abdel-Rehim:2015pwa} for the ensemble 
$cA2z.09.48$. As a setup strategy within the HMC, we produce an initial set of test vectors using three setup
iterations. Right before every inversion for $\rho_0$ we update the setup using one setup
iteration at $\rho_0$. The resulting preconditioner is used for all other Hasenbusch masses 
until the next update.}
\end{figure}

Empirically we find that the solver performance is stable for the tuned
parameter set. Within the ensemble $cA2.09.48$ we do not find any configurations 
where the iteration counts or the time to solution differ by more than 5\%, as can be seen in Figure~\ref{fig:observable}.
This behavior is also corroborated by the performance of the multigrid solver
during the force computation in the HMC algorithm,
where it shows very stable iteration counts for simulations at the physical point, see 
Figure~\ref{fig:hmc}.

\subsection{Analysis of parameter settings and tuning strategy} 
As demonstrated in the previous section, well-tuned parameters are important for good performance, and they tend 
to be stable at least for configurations from a given ensemble. It is therefore advisable to
invest some effort for obtaining good parameters. Since the parameter space is large,
it cannot be searched exhaustively, and there is thus a need for a strategy how to tune the method in practice   
 Our suggestion for twisted mass simulations is to keep the default parameters given in 
Table~\ref{tab:TMparams}, but tune the aggregation parameters and the twisted mass parameter on the
coarsest level, $\mu_{\rm coarse}$. This is justified by the analysis that we explain in the subsequent
sections as follows: 
In Section~\ref{sec:params_set1}, we present a novel analysis of the aggregation parameters without 
tuning $\mu_{\rm coarse}$, i.e.~we fix $\delta=1$; in Section~\ref{sec:trick}, we show the benefits obtained
by increasing $\mu_{\rm coarse}$ and we repeat the previous analysis; in Section~\ref{sec:trick4oMGs}, we demonstrate 
that also other multigrid approaches can benefit by an increasing $\mu_{\rm coarse}$.
All the tests are performed on one configuration averaging the time to solution for the u- and d-propagator.
This choice is motivated by the stability of the solver presented in the previous section.
\newpage
\subsubsection{Aggregation parameters\label{sec:params_set1}}

Aggregation parameters are the number of 
the test vectors $N_{v,\ell}$ and the size
of the lattice-blocks $V_{b,\ell}$ on each level $\ell$.  They should be tuned
simultaneously since they define the size of the coarser Dirac operator
and consequently the size of the projected subspace. In the present analysis, the solver is restricted to a 
3-level implementation. We do not find an improvement in the time to solution 
by using a 4th level, which is also the result found in Ref.~\cite{Frommer:2013kla} for the Wilson operator with similar 
lattice sizes. On the other hand, the inversion time increases when just a 2-level multigrid method is used. 

\begin{figure}
\includegraphics[width=1.0\textwidth]{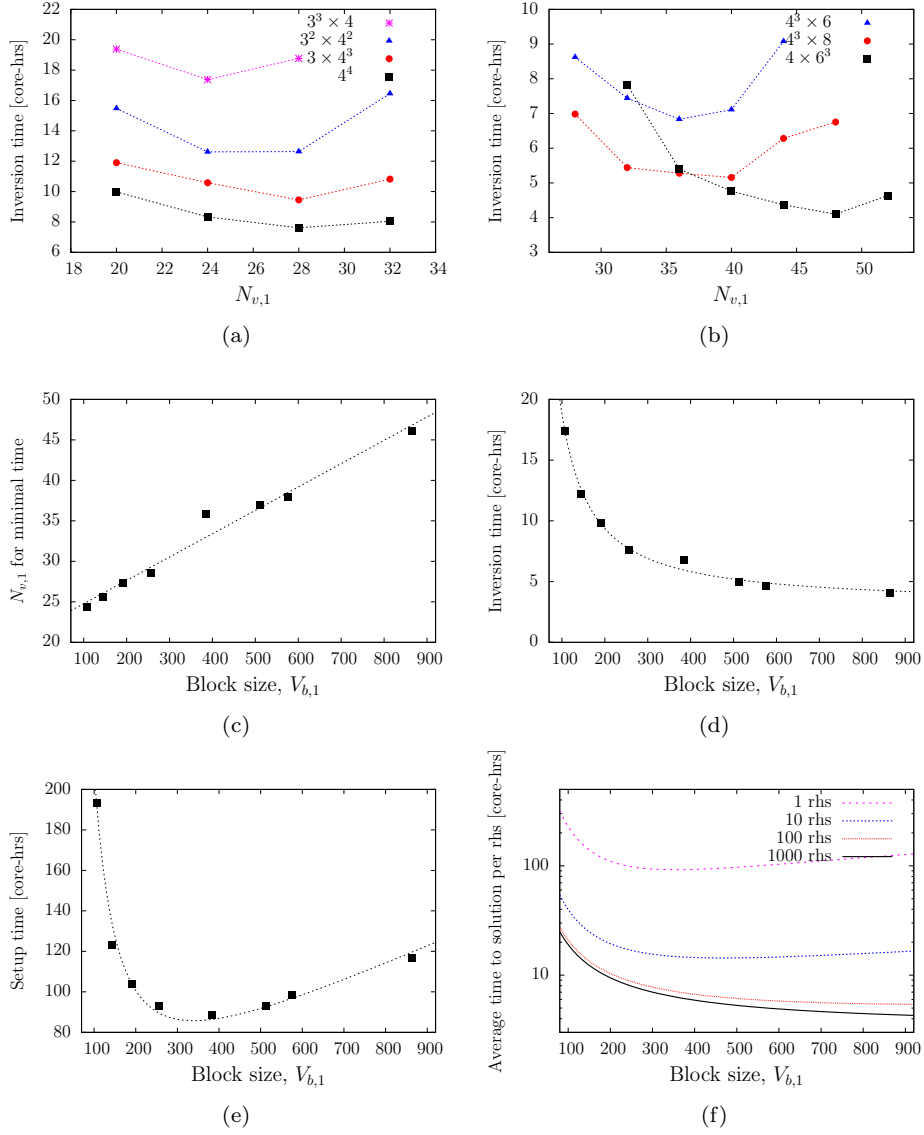}
\caption{\label{fig:blk_tv} Analysis of the aggregation parameters for the $cA2.09.48$ ensemble. For each set of
raw data  presented in (a) and (b)
the position and value of the minimum has been extracted with a parabolic interpolation and 
displayed in (c) and (d), respectively.
In (d), (e) and (f),
$N_{v,1}$ and $V_{b,1}$ are connected according to the minima 
found in (c), i.e.~(d) shows the
inversion time and (e) the setup time, 
both for the minima from (c).
The estimated average total time per right hand side 
$(t_{\setup}+n_{rhs}t_{solve})\big/ n_{rhs}$ is shown in (f).
The fitting functions are explained in the text.}
\end{figure}

To optimize the aggregation parameters on the first level, we fix the values of the parameters on the second level for which 
we find a block size of $V_{b,2} = 2^4$ and a number of test vectors around $28$ to work well. 
Fixing these parameters by setting $N_{v,2}=\max(28,N_{v,1})$ we analyze how the time needed to solve for one rhs 
and for the setup depends on the block size $V_{b,1}$ and the number of test vectors $N_{v,1}$.
The results are depicted in Figure~\ref{fig:blk_tv}, for the  $cA2.09.48$ ensemble and 
a scaling parameter $\delta=1$ on all levels has been used.

We find that every block size has an optimal number of test vectors as
shown in Figures~\ref{fig:blk_tv}(a) and \ref{fig:blk_tv}(b) that minimizes the time to solution.
By fitting the data to a polynomial of order 2 in $N_{v,1}$ we estimate the minima and find that the optimal number of 
test vectors grows approximately linearly for block sizes $V_{b,1}<8\cdot4^3$ with the block volume
\begin{equation}\label{eq:lindep_NV2}
N_{v,1}=\alpha +\beta\,V_{b,1},
\end{equation}
which is shown in Figure~\ref{fig:blk_tv}(c). This indicates a non-trivial connection between the 
fine grid size and the dimension of the coarse grid operator given by
\begin{equation}\label{eq:linNV}
 \textrm{dim}(D_{c,1}) = 2 \left(\frac{\alpha}{V_{b,1}} + \beta \right)V\,.
\end{equation}
This implies that the optimal size of the coarse grid operator increases linearly with the volume $V$ at 
fixed aggregation block size. For larger block sizes the behavior deviates from this linear
dependence, however. For this case  the minimum of the time to solution is already reached
for a smaller number of test vectors. We can interpret this complex behavior as giving some
insight into the non-trivial link between coarse grid size and local coherence.

The time to solution of the multigrid method is dominated by the solves with 
the coarse grid operator. We use a K-cycle and a 3-level approach
where, due to fixing the blocks of the coarsest grid, it follows that
$\textrm{dim}(D_{c,1}) \propto \textrm{dim}(D_{c,2})$. The numerical effort needed for a matrix-vector multiplication 
involving $D_{c,1}$ depends linearly on $\textrm{dim}(D_{c,1})$. Allowing for an
additional second order term we model the time for an inversion as
\begin{equation}
 t_{solve} = \gamma' + \delta' \, \textrm{dim}(D_{c,1}) + \varepsilon' \,\textrm{dim}(D_{c,1})^2.
\end{equation}
By using the dependence on $V_{b,1}$ of Eq.~\eqref{eq:linNV}
we can rewrite $t_{solve}$ with
\begin{equation}\label{eq:tsolve}
 t_{solve} = \gamma + \delta \, V_{b,1}^{-1} + \varepsilon \, V_{b,1}^{-2}.
\end{equation}
Within this model, connecting $N_{v,1}$ and $V_{b,1}$ according to Eq.~\eqref{eq:lindep_NV2} and 
Figure \ref{fig:blk_tv}(c), the time to solution can be fitted very well up to $V_{b,1} \approx 8^2\times4^2$
as shown in Figure~\ref{fig:blk_tv}(d).
\begin{figure}
\includegraphics[width=\textwidth]{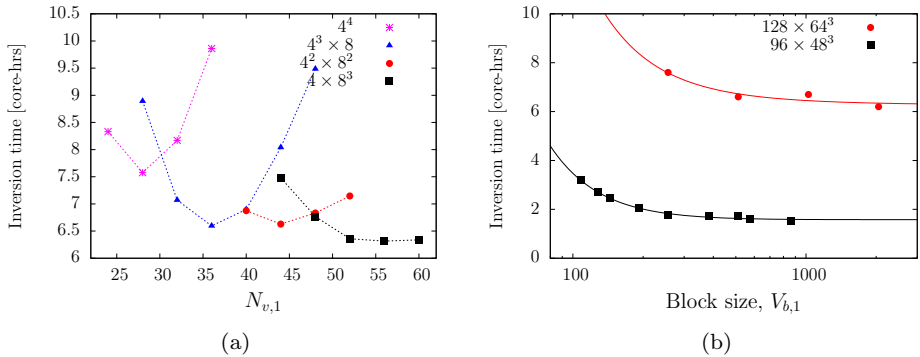}
\caption{\label{fig:big_blk} Analysis of aggregation parameters for $cA2.09.64$ and $\delta=5$ on the coarsest grid. 
The sets of raw data are presented in (a); from each set, the value of the minimum 
has been extracted with a parabolic interpolation and plotted with red points in (b).
The black points are for the ensemble  $cA2.09.48$ and $\delta=5$ on the coarsest grid.}
\end{figure}
In Figure~\ref{fig:big_blk}, we display the data obtained for the ensemble $cA2.09.64$ with a lattice size of $128\times64^3$.
The other lattice parameters are the same as those for $cA2.09.48$. In contrast to the discussed analysis of $cA2.09.48$, the coarse grid scale 
factor is set to $\delta=5$ on the coarsest grid. The full analysis on the dependence of the algorithm on $\delta$ is reserved for the next section.
For both lattice sizes the inversion times reach the minimum for the same block volumes.
The asymptotes $\gamma$ from Eq.~\eqref{eq:tsolve} are given by $1.64(13)$ for the $cA2.09.48$ ensemble and by $6.3(5)$ 
for the $cA2.09.64$ ensemble,  which is an increase of  $V^{5/4}$.
As mentioned above, for larger block sizes, the linear connection of $N_{v,1}$ and $V_{b,1}$ does no hold and the
solution time is increased. This is also observed for the time spent in the setup.
We also observe that with increasing lattice volume the region where the linear dependence holds is shifted.

During the setup procedure the solver is applied on $N_{v,1}$ test vectors 
for several  iterations. Here, the test vectors are orthonormalized at the end of each iteration
and used for building the multigrid hierarchy. We model the setup time by allowing for a linear and quadratic term as
\begin{equation}
t_{\setup} = \zeta + \eta\,N_{v,1}\,t_{solve}+ \theta\,\textrm{dim}(D_{c,1}),
\label{timings}
\end{equation}
where the amount of computation which does not involve the 
solver scales with $\textrm{dim}(D_{c,1})$ at leading order.
Figure~\ref{fig:blk_tv}(e) shows that the measured timings are indeed very well described by the Ansatz given in Eq.~\eqref{timings}. 
We  remark that for block sizes $V_b \sim 4^4$ up to $6^2 \times 4^2$ the time to solution shows a 
relatively large plateau, which makes the timings for the multigrid solver relatively stable. 
Further fine tuning in this region would lead only to small improvements.

The optimal choice for $N_{v,1}$ and $V_{b,1}$ depends on the number of rhs $n_{rhs}$. 
The total time is given by
\begin{equation}
t_{total} = t_{\setup} + n_{rhs}\,t_{solve},
\label{eq:ttotal}
\end{equation}
and we  find a non-trivial dependence of $V_{b,1}$ on $n_{rhs}$ for an optimal time to solution, 
as depicted in Figure~\ref{fig:blk_tv}(f). This motivated our suggestion to consider 
two different values for $n_{\setup}$, depending on the number of rhs.
\par

\subsubsection{Tuning the coarse grid scale factor $\delta$ \label{sec:trick}}

\begin{table}
\resizebox{\textwidth}{!}{
\centering
\begin{tabular}{l c c c c}
	\toprule
	Block size                 & Test vectors & $\delta$ for the coarsest $\mu$         & Setup time & Inversion time \\
	$V_{b,1}$                  & $N_{v,1}$    & $\mu_{\rm coarse}=\delta\mu$ & [core-hrs] & [core-hrs]     \\ \midrule
	$3^3\times\,4$             & 24           & 7.8                    & 31.1       & 3.19(3)        \\ \midrule
	$2\phantom{^1}\times\,4^3$ & 24           & 6.6                    & 27.5       & 2.71(5)        \\ \midrule
	$3^2\times\,4^2$           & 24           & 8.6                    & 23.5       & 2.47(2)        \\ \midrule
	$3\phantom{^1}\times\,4^3$ & 28           & 5.4                    & 28.0       & 2.04(5)        \\ \midrule
	$4^4$                      & 28           & 5.2                    & 22.2       & 1.75(4)        \\ \midrule
	$4^3\times\,6$             & 36           & 4.5                    & 37.4       & 1.74(6)        \\ \midrule
	$4^3\times\,8$             & 40           & 4.0                    & 40.7       & 1.73(5)        \\ \midrule
	$4^2\times\,6^2$           & 40           & 4.1                    & 37.7       & 1.59(6)        \\ \midrule
	$4\phantom{^1}\times\,6^3$ & 44           & 4.0                    & 38.3       & 1.52(4)        \\ \bottomrule
\end{tabular} }
\caption{\label{tab:blk_cmu}Summary of $\delta$ parameters 
yielding the optimal solve time for various block sizes $V_{b,1}$. 
The shown numbers were computed for the $cA2.09.48$ ensemble and a 
relative residual tolerance of $10^{-9}$. The number of test vectors
was chosen as in Figure \ref{fig:blk_tv}(c) according to the block size $V_{b,1}$. 
All other parameters were fixed to the values in Table~\ref{tab:TMparams}.}
\end{table}

\begin{figure}
\includegraphics[width=\textwidth]{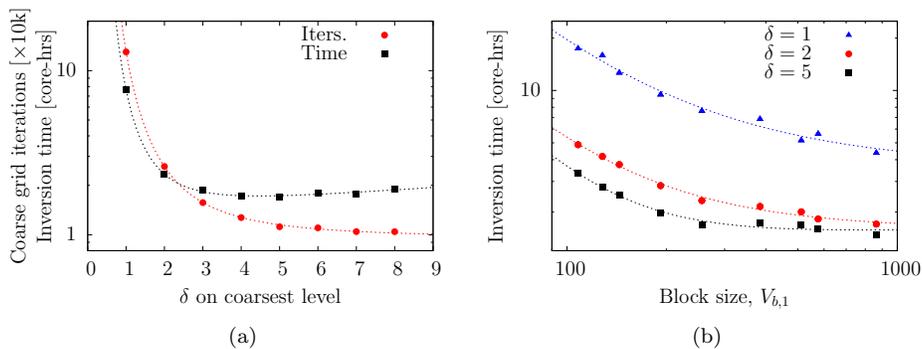}
\caption{\label{fig:blk_cmu}  The number of coarse grid iterations as a function of the $\delta$ parameter and the inversion
time for different $\delta$-values depending on the block size. The results
depicted in (a) are for a block size of $V_{b,1}=4^4$ 
and compare the number of iterations on the coarse grid with the fine grid inversion time.
The number of iterations has been scaled down by a factor $10\,000$. The behavior for
the coarse grid iterations is proportional to $1/\delta^2$, 
while the inversion time has an additional term linear in $\delta$ 
due to the increase in the fine grid iteration count. 
The results depicted in (b) show the inversion time for 
different values of $\delta$ and different block sizes. The behavior is similar  
to the one observed in Figure~\ref{fig:blk_tv}(d). }
\end{figure}

At physical quark masses, the density of the low-lying eigenvalues for the
twisted mass operator increases compared to the Wilson Dirac operator, as explained in Section~\ref{sec:Wtwisted}. 
Densely populated low eigenvalues slow down the iteration of the Krylov subspace solvers on 
the coarsest grid and thus of the whole multigrid method. This is much more pronounced for the twisted mass Dirac
operator than it is for the Wilson Dirac operator. For the latter, Table~\ref{tab:clover} reports 
a total of around $3\,000$ coarse grid iterations for the considered ensemble, whereas data depicted 
in Figure~\ref{fig:blk_cmu}(a) show that for $\delta = 1$ we have roughly $40$ times more 
coarse grid iterations in the twisted mass case, an unexpectedly large increase. 

When using a larger twisted
mass value on the coarsest gird operator, given by $\mu_{\rm coarse} = \delta \mu$ with $\delta \geq 1$, we
make the small eigenvalues less dense in the low-lying part of the spectrum. This
speeds up the inversion time on the coarsest grid. We also analyze the effect of the 
scaling factor $\delta$ as a function of the block size. The results are depicted in Figure~\ref{fig:blk_cmu} 
and we  find that the optimal aggregation parameters determined in the previous section
do not depend on the chosen coarse grid scale factor $\delta$.
In Table~\ref{tab:blk_cmu} we present the final results for the $cA2.09.48$ ensemble,
summarizing the $\delta$ parameters which minimize the inversion time for various 
block sizes and the corresponding optimal numbers of test vectors.
By using a scaling factor $\delta = 2$ the iteration count on the coarsest grid 
is already reduced by a factor of five, which results into 
a speed-up of the fine grid inversion time by roughly a factor of three.
Optimal performance is achieved for a relatively large plateau around $\delta \sim 5$.  
Obviously, a large scaling factor $\delta$ causes a distinct violation of the Galerkin
condition, i.e.~$P^\dagger D(\mu) P \neq D_{c}(\delta\mu)$. However, we only find a minor
increase in the iteration count of the fine grid solver from 14 to 16 iterations.

\subsubsection{The scale factor $\delta$ and different multigrid approaches}\label{sec:trick4oMGs}

In addition to the studies of the delta parameter in DD-$\alpha$AMG we tested the influence of 
a delta parameter in the QUDA implementation of MG-GCR \cite{QUDA:quda} and in openQCD
\footnote{We use for the tests a modified openQCD-version, which is optimized for twisted mass fermions, i.e.~where the even--odd
reduced twisted mass Dirac operator is implemented.}~\cite{Luscher:OpenQCD}. 
The latter is restricted to a two-level approach. For both approaches we found that the iteration 
counts on the coarsest grid were reduced by using a scale factor $\delta > 1$, also resulting in reduced 
inversion and setup times. For $\delta \approx 5$ we observe speed-ups similar to those for DD-$\alpha$AMG 
for the $cA2.09.48$ ensemble.

For MG-GCR this behavior is to be expected since both approaches share the same type of $\Gamma_5$-respecting
interpolation (cf.~Eq.~\eqref{eq:gamma5_symmetry}). However, the interpolation within the openQCD multigrid solver 
is not $\Gamma_5$-respecting, resulting in a non-diagonal summand $P^\dagger \Gamma_5 P \delta \mu$ 
on the coarse grid instead of $\Gamma_{5,c} \delta \mu$ in Eq.~\eqref{delta:eq}.

We indeed find cases where the openQCD solver and DD-$\alpha$AMG show different behaviors. 
When going to the $cA2.09.64$ ensemble with a bigger volume, we observe that the optimal $\delta$
for the openQCD solver increases while it remains constant for DD-$\alpha$AMG.
Furthermore, when solving for $D(-\mu)$ with a setup built for $D(\mu)$ from the $cA2.09.48$
ensemble, we find that the optimal $\delta$ increases by a factor of four within the openQCD 
solver whereas it remains constant for DD-$\alpha$AMG.

For the Wilson fermion case it was reported in Ref.~\cite{Frommer:2013kla} that the two-level openQCD 
solver shows about the same performance as the three-level DD-$\alpha$AMG approach. 
For the twisted mass ensemble $cA2.09.48$, the openQCD solver is roughly a factor 
of four slower than DD-$\alpha$AMG.

\section{Conclusions and outlook\label{sec:conclusions}}
The DD-$\alpha$AMG solver is extended to the case of $N_f=2$ twisted mass fermions. By carefully tuning the 
parameters we show that one can achieve speed-ups similar to those obtained
in the case of Wilson fermions, to be precise, a factor of 200 compared to CG.
This is mainly achieved by adapting the twisted mass 
parameter on the coarsest grid operator such that computing the coarsest grid correction
becomes less time-consuming.
Using a factor $\delta\sim5$ to increase $\mu_{\rm coarse}=\delta\cdot\mu$ 
on the coarsest level decreases the coarse grid iteration count by
a roughly factor around 10 and improves the time to solution by a factor of around 4
for the case of ensembles generated at the physical pion mass.

An optimal set of parameters for different applications of the solver is presented
and the strategy for tuning the aggregation parameters and the 
factor $\delta$ for the twisted mass parameter is discussed in detail. 
Using the optimal set of parameters, inversions are more  
than two orders of magnitude faster as compared to standard CG.
Within the HMC, DD-$\alpha$AMG achieves a speed-up of an order of magnitude compared to standard CG.

The DDalphaAMG library for twisted mass fermions is publicly available, 
and an interface to the tmLQCD software package is provided.
The technical details are summarized in the Appendix~\ref{sec:library}.

For the future we plan to extend multigrid methods to
the heavy doublet sector of the twisted mass formulation.

\subsection*{Acknowledgments}
This project has received funding from the  Horizon 2020 research and innovation programme of the European Commission
under the Marie Sklodowska-Curie grant agreement No 642069. S.B.~is supported by this programme.
We would like to thank Giannis Koutsou for fruitful  discussions and for helping us accessing the configurations.
We also thank Artur Strebel and Simon Heybrock for guidance with the DD-$\alpha$AMG code and Bj\"orn
Leder for his suggestion to shift the twisted mass on the coarse grid
to speed-up the coarse grid solver. The authors gratefully acknowledge 
the Gauss Centre for Supercomputing e.V.~for funding the project \emph{pr74yo} by 
providing computing time on the GCS Supercomputer SuperMUC at Leibniz Supercomputing Centre,
the computing time granted by the John von Neumann Institute for Computing 
(NIC) and provided on the supercomputer JURECA at Jülich Supercomputing Centre (JSC) through the grant \emph{ecy00}, the High Performance Computing Center in Stuttgart for providing computation time on the High
Performance Computing system Hazel Hen for calculating the eigenvalues through the grant \textit{GCS-Nops
(44066)}, and the computational resources on Cy-Tera at the Cyprus Supercomputing Centers through the grant \emph{lspre258s1}.

\appendix
\section{Overview of the DDalphaAMG library\label{sec:library}}
The DDalphaAMG solver library, available at Ref.~\cite{Rottmann:DDalphaAMG},
has been recently released under GNU General Public License.
This software package includes an implementation of the DD-$\alpha$AMG
for clover Wilson fermions as described in Ref.~\cite{Frommer:2013fsa}.
The implementation is of production code quality, it includes
a hybrid MPI/\-openMP parallelization, state-of-the-art mixed precision and
odd-even preconditioning approaches and also SSE3 optimizations. 
Implementation details can be found in Ref.~\cite{RottmannPhD}. 

Based on the DDalphaAMG code we developed a version, which supports twisted mass fermions,
available at Ref.~\cite{Bacchio:DDalphaAMG}. We added the following features to the library:
$N_f=2$ twisted mass fermions and twisted boundary conditions are supported, and different twisted mass 
shifts on the even and odd sites can be applied, 
which are required for the Hasenbusch mass preconditioning in the HMC when even-odd
preconditioning is used.

An user documentation for the library can be found in \texttt{src/DDalphaAMG.h}, and a sample code
that illustrates how to use the library interface functions can be found in~\texttt{tests/DDalphaAMG\_sample.c}.
Moreover, the library has been integrated in the tmLQCD program~\cite{Jansen:2009xp},
therefore we also adjusted the library interface. The interfaced code is available at Ref.~\cite{Finkenrath:tmLQCD}. Details on the employment of DDalphaAMG within tmLQCD are available at Ref.~\cite{Bacchio:2016bwn}.
  
\bibliography{common}

\end{document}